\documentclass[3p,times]{elsarticle}

\usepackage{ecrc}

\volume{00}

\firstpage{1}

\journalname{..}

\runauth{}

\usepackage{amssymb}
\usepackage{amsthm}

\usepackage{multirow}

\usepackage[figuresright]{rotating}

\begin{document}

\begin{frontmatter}



\dochead{}

\title{Measurement of Event Plane Correlations in Pb-Pb Collisions at $\sqrt{s_{\mathrm{NN}}}$=2.76 TeV with the ATLAS Detector}


\author{Jiangyong Jia on behalf of the ATLAS Collaboration}

\address{Chemistry Department, Stony Brook University, Stony Brook, NY 11794, USA}
\address{Physics Department, Brookhaven National Laboratory, Upton, NY 11796, USA}

\begin{abstract}
A measurement of correlations between event-plane angles $\Phi_n$ is presented as a function of centrality for Pb-Pb collisions at $\sqrt{s_{_{\mathrm{NN}}}}=2.76$ TeV. These correlations are estimated from observed event-plane angles $\Psi_n$ obtained from charged particle or transverse energy flow measured over a large pseudorapidity range $|\eta|<4.8$, followed by a resolution correction that accounts for the dispersion of $\Psi_n$ relative to $\Phi_n$. Various correlators involving two or three event planes with acceptable resolution are measured. Significant positive correlations are observed for $4(\Phi_2-\Phi_4)$, $6(\Phi_2-\Phi_6)$, $6(\Phi_3-\Phi_6)$, $2\Phi_2+3\Phi_3-5\Phi_5$, $2\Phi_2+4\Phi_4-6\Phi_6$ and $-10\Phi_2+4\Phi_4+6\Phi_6$. However, the measured correlations for $2\Phi_2-6\Phi_3+4\Phi_4$ are negative. These results may shed light on the patterns of the fluctuation of the created matter in the initial state as well as the subsequent hydrodynamic evolution.
\end{abstract}


\end{frontmatter}







\bibliographystyle{elsarticle-num}


In recent years, the measurement of harmonic flow coefficients $v_n$ has provided important insight into the hot and dense matter created in heavy ion collisions at the Relativistic Heavy Ion Collider (RHIC) and the Large Hadron Collider (LHC). These coefficients are generally obtained from a Fourier expansion of particle production in azimuthal angle $\phi$, $\frac{dN}{d\phi}\propto1+2\sum_{n=1}^{\infty}v_{n}\cos n(\phi-\Phi_{n})$, where $\Phi_n$, known as the event plane (EP), represents the phase of $v_n$. Experimentally, $\Phi_n$ is estimated using the azimuthal distribution of particles in the event. The estimated EP direction $\Psi_n$ is referred to as the observed event plane, to distinguish it from $\Phi_n$. 

Besides the $v_2$ coefficient, which reflects mainly the hydrodynamic response of the matter to the elliptic shape of the overlap region, significant first order ($v_1$) and higher order ($v_3$--$v_6$) coefficients were also obtained~\cite{Aamodt:2011by,Aad:2012bu,CMS:2012wg}. These non-elliptic harmonics suggest that the matter densities can fluctuate in the initial state, and the harmonics simply reflects the hydrodynamic responses to these fluctuations. More details about these fluctuations can be inferred from the correlations between $\Phi_{n}$ of different order. Calculations based on Glauber model suggest the existence of significant correlations in the initial state~\cite{Teaney:2010vd,Nagle:2010zk,Bhalerao:2011yg,Jia:2012ma}; the correlations can also be generated dynamically in hydrodynamic evolution due to non-linear effects~\cite{Qiu:2011iv}. This proceedings present measurement of EP correlations~\cite{Rpcorr}, which can shed light on these initial state fluctuations and dynamic mixing between different $\Phi_n$ in the final state.

The correlation between several EPs are characterised by a set of cosine functions, $\langle\cos (c_1\Phi_{1}+...+lc_l\Phi_{l})\rangle$ with $c_1+2c_2...+lc_l=0$~\cite{Bhalerao:2011yg}. Two-plane correlators can be written as $\langle\cos k(\Phi_{n}-\Phi_{m})\rangle$, where $k$ is the least common multiple of $n$ and $m$. Experimentally, these correlations are obtained from the observed event planes as~\cite{Jia:2012ma}:
\begin{eqnarray} 
\label{eq:o3a}
\langle\cos (c_1\Phi_{1}+...+lc_l\Phi_{l})\rangle = \frac{\langle\cos (c_1\Psi_{1}+...+lc_l\Psi_{l})\rangle} {\mathrm{Res}\{c_1\Psi_1\}...\mathrm{Res}\{c_ll\Psi_l\}},\;\;\mathrm{Res}\{c_nn\Psi_n\} = \langle\cos c_nn(\Psi_n-\Phi_n)\rangle
\end{eqnarray}
where the resolution factors $\mathrm{Res}\{c_nn\Psi_n\}$ are determined using the standard two-subevent (2SE) or three-subevent (3SE) methods~\cite{Poskanzer:1998yz}. To avoid auto-correlations, the $\Psi_n$ needs to be measured using subevents covering different $\eta$ ranges, preferably with a gap in between. A large number of correlators could be studied. However, the measurability of these correlators is dictated by the value of ${\mathrm{Res}\{jn\Psi_{n}\}}$; they are measurable in ATLAS for $n=2$--6 and $j$ values up to 6~\cite{Rpcorr}. This analysis focuses on two- and three-plane correlations. The latter can be generally constructed from several two-plane relative angles, and hence probes the correlations of two angles relative to the third~\cite{Jia:2012ma}, such as:
\begin{eqnarray}
\nonumber
&&2\Phi_2+4\Phi_4-6\Phi_6   = 4(\Phi_4-\Phi_2)-6(\Phi_6-\Phi_2),\;\;-10\Phi_2+4\Phi_4+6\Phi_6 = 4(\Phi_4-\Phi_2)+6(\Phi_6-\Phi_2)\\\label{eq:o2b}
&&2\langle\sin4(\Phi_{4}-\Phi_{2})\sin6(\Phi_{6}-\Phi_{2})\rangle = \langle\cos(2\Phi_{2}+4\Phi_{4}-6\Phi_{6})\rangle-\langle\cos(-10\Phi_{2}+4\Phi_{4}+6\Phi_{6})\rangle
\end{eqnarray}

In a two-plane correlation (2PC), the event is divided into two subevents symmetrically around $\eta=0$ with a gap in between such that nominally they have the same resolution. Each subevent provides its own measurement of the two event planes: $\Psi_n^{P}$ and $\Psi_m^{P}$ for positive $\eta$ and $\Psi_n^{N}$ and $\Psi_m^{N}$ for negative $\eta$. This leads to two statistically independent measurements of the correlator, which are averaged to obtain the final signal:
\begin{eqnarray}
\label{eq:2db}
 \langle\cos k(\Phi_n-\Phi_m)\rangle= \frac{\left\langle\cos k(\Psi_n^P-\Psi_m^N)\right\rangle+\left\langle\cos k(\Psi_n^N-\Psi_m^P)\right\rangle}{\mathrm{Res}\{k\Psi_n^{\mathrm{P}}\}\mathrm{Res}\{k\Psi_m^{\mathrm{N}}\}+\mathrm{Res}\{k\Psi_n^{\mathrm{N}}\}\mathrm{Res}\{k\Psi_m^{\mathrm{P}}\}},
\end{eqnarray}

In a three-plane correlation (3PC), three non-overlapping subevents, labeled as A, B and C, are chosen to have similar $\eta$ coverage. In this analysis, subevent A and C are chosen to be symmetric about $\eta=0$, and hence have identical resolution, while the resolution of B can be different. The symmetry between A and C reduces the 6 independent combinations into three pairs of measurements, labeled as Type1, Type2 and Type3. For example, the Type1 measurement of the correlation $2\Phi_2+3\Phi_3-5\Phi_5$ is obtained by requiring the $\Psi_2$ plane to be measured by subevent B:
\begin{eqnarray}
\label{eq:2e}
\left\langle\cos (2\Phi_2+3\Phi_3-5\Phi_5)\right\rangle_{\mathrm{Type1}}=\frac{\left\langle\cos (2\Psi_2^B+3\Psi_3^A-5\Psi_5^C)\right\rangle+\left\langle\cos  (2\Psi_2^B+3\Psi_3^C-5\Psi_5^A)\right\rangle} {\mathrm{Res}\{2\Psi_2^{B}\}\mathrm{Res}\{3\Psi_3^{A}\}\mathrm{Res}\{5\Psi_5^{C}\}+\mathrm{Res}\{2\Psi_2^{B}\}\mathrm{Res}\{3\Psi_3^{C}\}\mathrm{Res}\{5\Psi_5^{A}\}},
\end{eqnarray}
Similarly, the Type2 (Type3) measurement is obtained by requiring the $\Psi_3$ ($\Psi_5$) to be measured by subevent B. 

This analysis is based on approximately 8~$\mu\mathrm{b}^{-1}$ of minimum bias Pb-Pb data collected in 2010 at $\sqrt{s_{_{\mathrm{NN}}}}=2.76$~TeV~\cite{Aad:2012bu}. Three subsystems of ATLAS~\cite{Aad:2008zzm} are used to measure the event plane: the inner detector (ID) over $|\eta|<2.5$, the barrel and endcap electromagnetic calorimeters (ECal) covering $|\eta|<3.2$ and the forward calorimeter (FCal), which extend the calorimeter coverage to $|\eta|<4.9$. These detectors are divided into a set of segments each covering 0.5 to 0.7 units in $\eta$, and the subevents are constructed by combining these segments. The subevent combinations for the two- and three-plane correlations are listed in Table~\ref{tab:dets}. The combination of ECal and FCal (ECalFCal) provides the default measurement, while the ID serves as cross-checks.
\begin{table}[h]
\centering
\begin{tabular}{|c|c||r|r|r|r|r|r|}\hline 
\multirow{2}{*}{Two-plane correlation (A,B)} &Default     &\multicolumn{3}{|r|}{\hspace{0.8cm}ECalFCal$_{P}$ $\eta\in$(0.5,4.8)} & \multicolumn{3}{|r|}{ECalFCal$_{N}$ $\eta\in$(-4.8,-0.5)}\tabularnewline\cline{2-8} 
&Cross-check &\multicolumn{3}{|r|}{ID$_{P}$ $\eta\in$(0.5,2.5)}       & \multicolumn{3}{|r|}{ID$_{N}$ $\eta\in$(-2.5,-0.5)}       \tabularnewline\hline\hline
\multirow{2}{*}{Three-plane correlation (A,B,C)}& Default     &\multicolumn{2}{|r|}{ECal$_{P}$ $\eta\in$(0.5,2.7)}  & \multicolumn{2}{|r|}{FCal $|\eta|\in$(3.3,4.8)}   & \multicolumn{2}{|r|}{ECal$_{N}$ $\eta\in$(-2.7,-0.5)}\tabularnewline\cline{2-8}
&Cross-check &\multicolumn{2}{|r|}{ID$_{P}$ $\eta\in$(1.5,2.5) }   & \multicolumn{2}{|r|}{ID  $\eta\in$(-1.0,1.0)}         & \multicolumn{2}{|r|}{ID$_{N}$ $\eta\in$(-2.5,-1.5)}  \tabularnewline\hline
\end{tabular}\caption{\label{tab:dets} The definition of the subevents and their $\eta$ coverage using the two-plane and three-plane correlations.}
\end{table}

\begin{figure}[h!]
\begin{center}
\includegraphics[width=0.8\columnwidth]{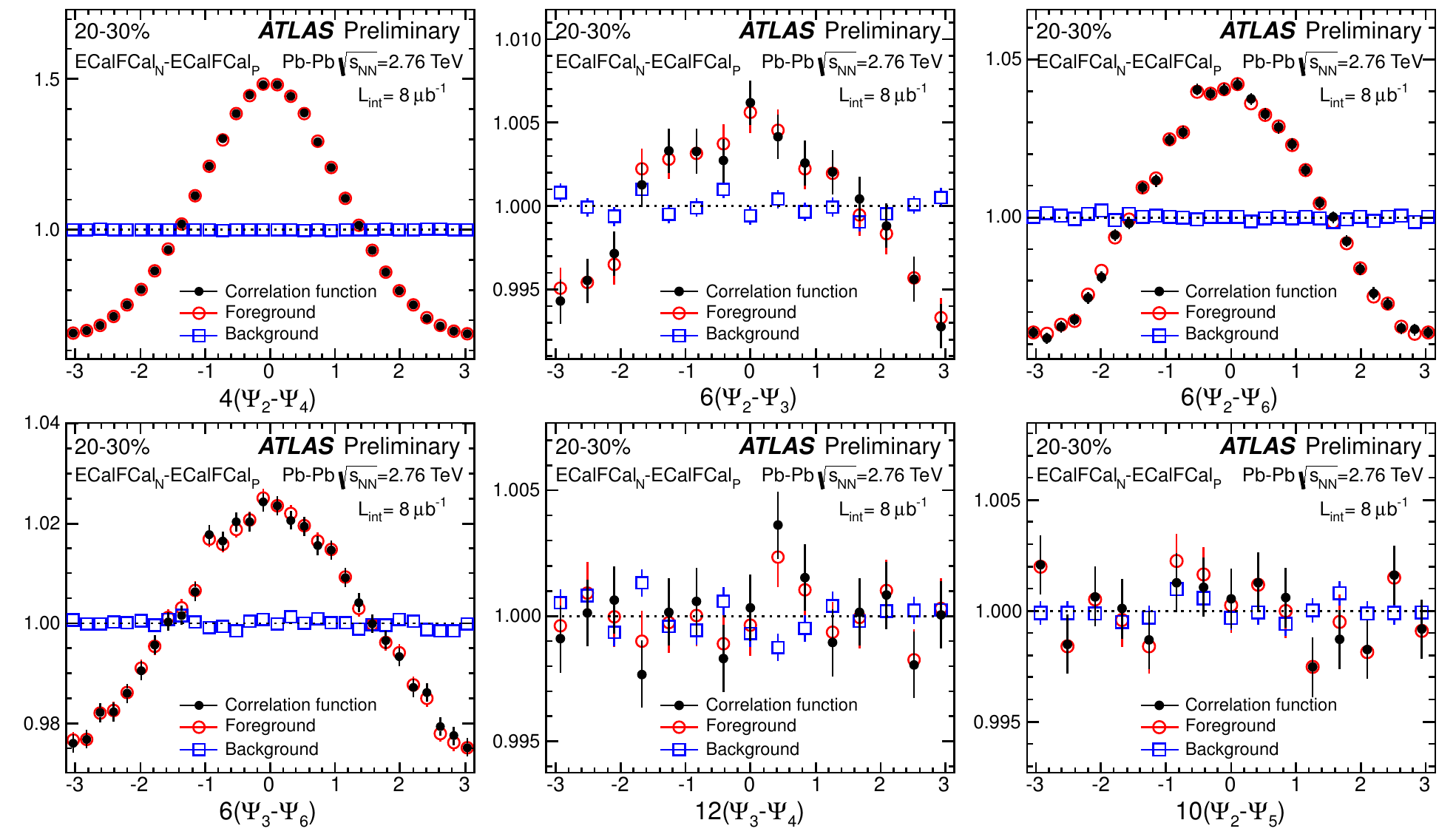}
\vspace*{-0.5cm}
\end{center}
\caption{\label{fig:2pcraw} Relative angle distributions between two event planes for the (20--30)\% centrality interval for foreground (open circles), background (open squares) and correlation function (filled circles). The correlation functions are used via Eq.~\ref{eq:2db} to obtain the two-plane correlators.}
\end{figure}

Figure~\ref{fig:2pcraw} shows the two-plane relative angle distributions for the (20--30)\% centrality interval from the FCalECal. To estimate the detector effects, a background distribution is calculated by combining $\Psi_{n}$ (or $\Psi_{m}$) in one event with $\Psi_{m}$ (or $\Psi_{n}$) in another event with similar centrality (matched within 5\%) and primary vertex $z$-position (matched within 3~cm). The correlation function is then obtained as the ratio of same event correlation (S) to the background distribution (B):
\begin{eqnarray}
\label{eq:cf1}
C(k(\Psi_{n}-\Psi_{m})) &=& \frac{S(k(\Psi_{n}-\Psi_{m}))}{B(k(\Psi_{n}-\Psi_{m}))}
\end{eqnarray}
The correlation functions show significant positive signals for $4(\Psi_2-\Psi_4)$, $6(\Psi_2-\Psi_3)$, $6(\Psi_2-\Psi_6)$ and $6(\Psi_3-\Psi_6)$. The observed correlation signals are calculated as the cosine averages of these correlation functions.

The resolution factors $\mathrm{Res}\{jn\Psi_{n}\}$ are determined using the 2SE and 3SE methods, with their differences as part of the systematic uncertainty. The potential influence of non-flow effects due to short-range correlations are studied by requiring a minimal $\eta$ gap, $\eta_{\mathrm{min}}$, between the two ECalFCal subevents. Results obtained for $\eta_{\mathrm{min}}$ values in the range of 0 to 6 are found to be consistent. In all cases, the influences of these short-range correlations are negligible for $\eta_{\mathrm{min}}>0.4$ (the choices of the subevents in Table~\ref{tab:dets} have a minimum $\eta$ gap of 0.6).

\begin{figure}[h!]
\begin{center}
\includegraphics[width=0.8\columnwidth]{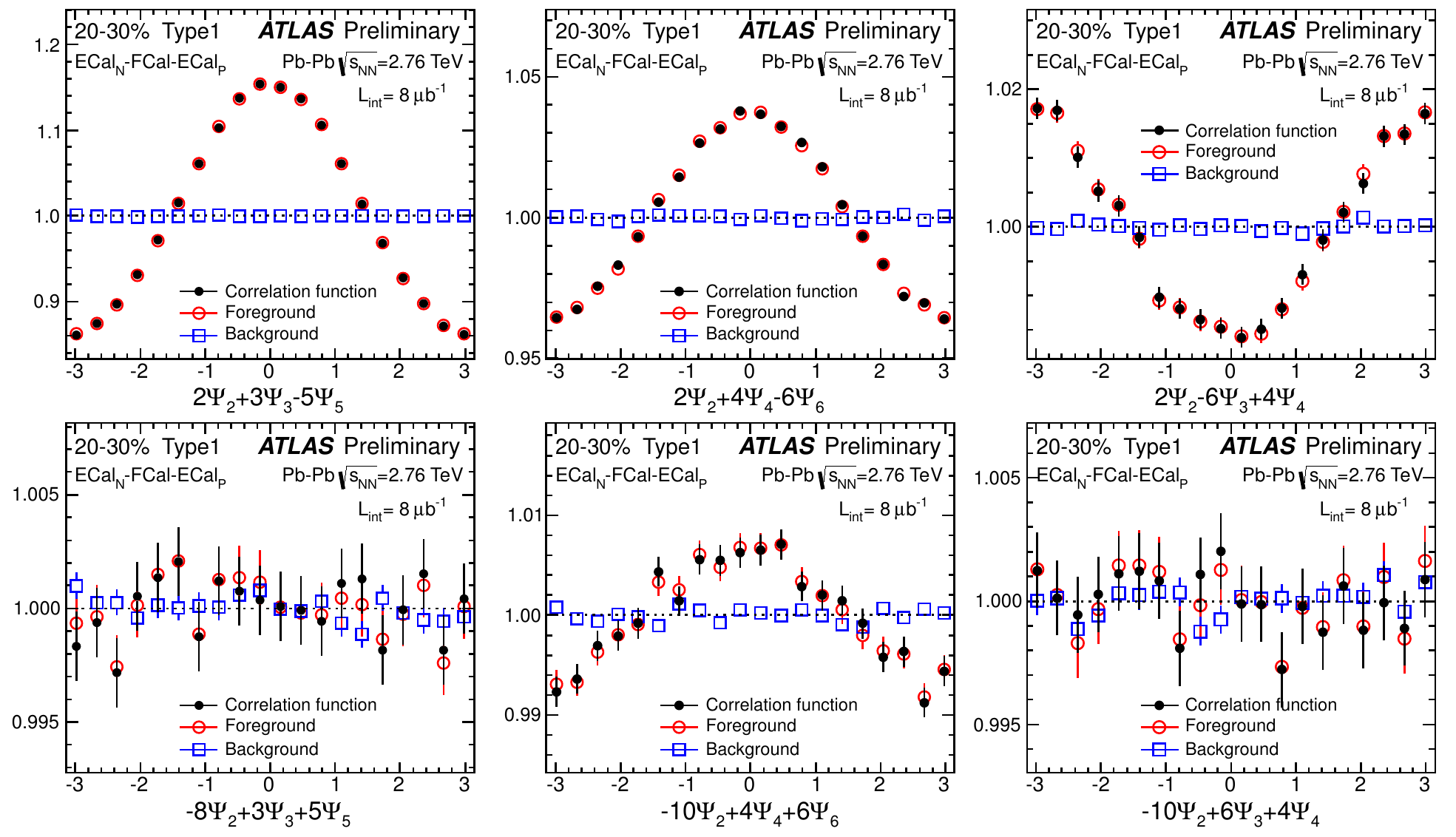}
\vspace*{-0.5cm}
\end{center}
\caption{\label{fig:3pcraw} Relative angle distributions between three event planes for the (20--30)\% centrality interval for Type1 correlation for foreground (open circles), background (open squares) and correlation function (filled circles). The correlation functions are used via equations similar to Eq.~\ref{eq:2e} to obtain the three-plane correlators.}
\end{figure}

The analysis procedure discussed above is also valid for the 3PC analysis, except that the three independent measurements of 3PC for each correlator need to be combined. Figure~\ref{fig:3pcraw} shows the relative angle distributions for various three-plane correlators from the Type1 measurement in the (20--30)\% centrality interval. The observed correlation signals are calculated as cosine average of the correlation function in a generalization of Eq.~\ref{eq:cf1}:
\begin{eqnarray}
\label{eq:cf2}
C(c_nn\Psi_{n}+c_mm\Psi_{m}+c_hh\Psi_{h}) &=& \frac{S(c_nn\Psi_{n}+c_mm\Psi_{m}+c_hh\Psi_{h})}{B(c_nn\Psi_{n}+c_mm\Psi_{m}+c_hh\Psi_{h})}
\end{eqnarray}
where the background distribution is constructed by requiring $\Psi_{n}$, $\Psi_{m}$, and $\Psi_{h}$ from different events. The correlation functions show significant positive signals for $2\Psi_{2}+3\Psi_{3}-5\Psi_5$,  $2\Psi_{2}+4\Psi_{4}-6\Psi_6$, and $-10\Psi_{2}+4\Psi_{4}+6\Psi_6$, while the signal for $2\Psi_{2}-6\Psi_{3}+4\Psi_4$ is negative, and the signals for the remaining correlators are consistent with zero. 

\begin{figure}[h!]
\begin{center}
\includegraphics[width=1\columnwidth]{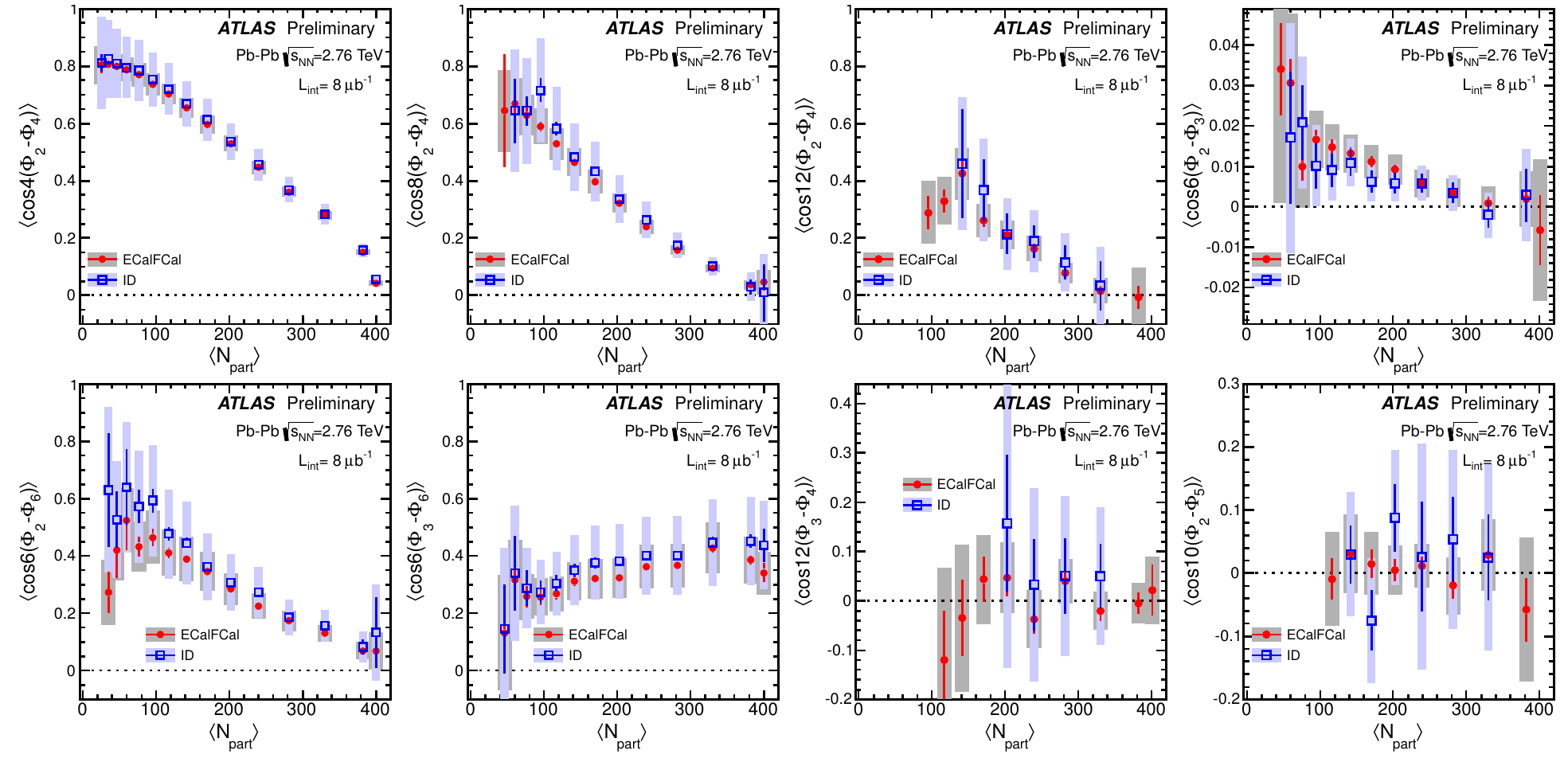}
\vspace*{-1cm}
\end{center}
\caption{\label{fig:result1} The final corrected eight two-plane correlators $\left\langle \cos jk(\Phi_n-\Phi_{m}) \right\rangle$ as a function of $\langle N_{\mathrm{part}}\rangle$ measured by ECalFCal and ID. The middle two panels in the top row have $j=2$ and $j=3$, respectively, while all other panels have $j=1$. The error bars and shaded bands indicate the statistical uncertainty and total systematic uncertainty, respectively.}
\end{figure}

\begin{figure}[h!]
\begin{center}
\includegraphics[width=1\columnwidth]{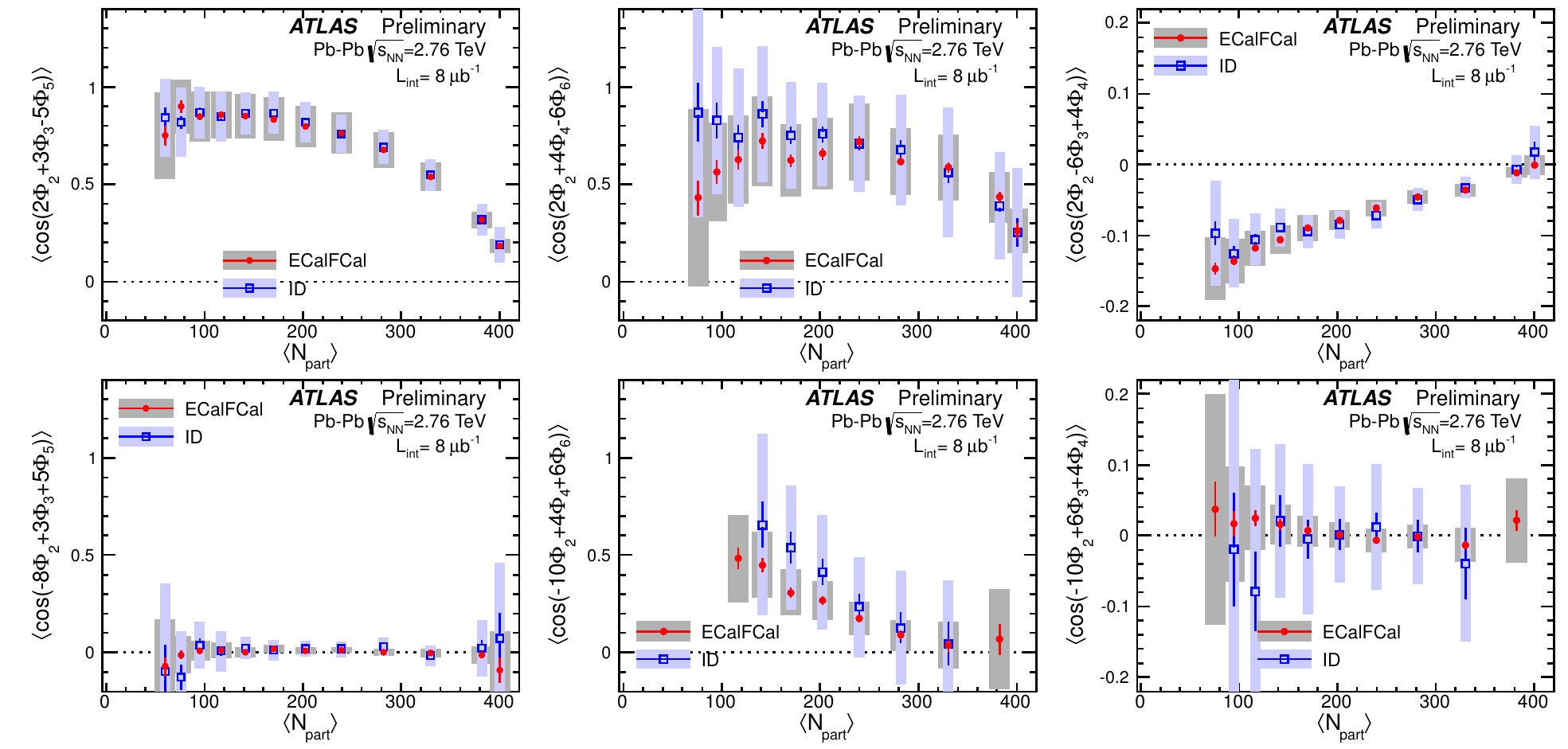}
\vspace*{-1cm}
\end{center}
\caption{\label{fig:result2} The final corrected six three-plane correlators as a function of $\langle N_{\mathrm{part}}\rangle$ measured by ECalFCal and ID. The error bars and shaded bands indicate the statistical uncertainty and total systematic uncertainty, respectively.}
\end{figure}

Figures~\ref{fig:result1} and \ref{fig:result2} show the centrality dependence, given here by average number of participanting nucleons $\langle N_{\mathrm{part}}\rangle$, of the two-plane and three-plane correlators, respectively. The systematic uncertainties are strongly correlated across the centrality range shown. Given that the ID is a different type of detector and measures only charged particles, the consistency between the ID and ECalFCal attests to the robustness of this measurement. Large positive values are observed in most cases and their magnitudes usually decrease with $\langle N_{\mathrm{part}}\rangle$. The value of  $\left\langle\cos6(\Phi_2-\Phi_3)\right\rangle$ is small ($<0.02$), yet exhibits a similar dependence on $\langle N_{\mathrm{part}}\rangle$. Interestingly, two other correlators show very different trends: the value of $\left\langle\cos6(\Phi_3-\Phi_6)\right\rangle$ increases with $\langle N_{\mathrm{part}}\rangle$, and the value of $\left\langle\cos(2\Phi_2-6\Phi_3+4\Phi_4)\right\rangle$ is negative and its magnitudes decreases with $\langle N_{\mathrm{part}}\rangle$. The centrality dependence trends for some correlators are qualitatively similar to the predictions of the Glauber model~\cite{Nagle:2010zk,Bhalerao:2011yg,Jia:2012ma}, such as $\left\langle\cos4(\Phi_2-\Phi_4)\right\rangle$, $\left\langle\cos6(\Phi_2-\Phi_6)\right\rangle$, $\left\langle\cos6(\Phi_2-\Phi_3)\right\rangle$, $\left\langle\cos(2\Phi_2+3\Phi_3-5\Phi_5)\right\rangle$ and $\left\langle\cos(2\Phi_2+4\Phi_4-6\Phi_6)\right\rangle$; however, the Glauber model fails to reproduce the magnitudes and centrality dependences for other correlators such as $\left\langle\cos6(\Phi_3-\Phi_6)\right\rangle$, $\left\langle\cos(2\Phi_2-6\Phi_3+4\Phi_4)\right\rangle$ and \\$\left\langle\cos(-10\Phi_2+4\Phi_4+6\Phi_6)\right\rangle$. This behavior might suggest the importance of non-linear response of the hydrodynamic evolution of the medium to the fluctuations in the initial geometry~\cite{Qiu:2011iv}.


In summary, measurements of several correlators between two and three event planes are presented using 8 $\mu$b$^{-1}$ Pb-Pb collision data at $\sqrt{s_{_{\mathrm{NN}}}}=2.76$ TeV~\cite{Rpcorr}. Several different trends in the centrality dependence of these correlators are observed. Some of the trends are qualitatively similar with predictions based on the Glauber model, while others differ significantly. These observations suggest that both the fluctuations in the initial geometry and dynamical evolution of the medium in the final state are important for creating these correlations in momentum space. A detailed theoretical description of these correlations may lead to new insights into the patterns of the fluctuation of the created matter in the initial state as well as the subsequent hydrodynamic evolution in heavy ion collisions.

This work is in part supported by NSF under award number PHY-1019387.

\end{document}